\documentclass[10pt]{article}
\usepackage{latexsym}
\usepackage{amssymb}
\usepackage{amsfonts}
\usepackage{amsmath}
\usepackage{theorem}

\theorembodyfont{\upshape}

\newtheorem{theorem}{Theorem}[section]

\begin{document}

\title{\bf On the ``Matrix Approach'' to Interacting Particle Systems} 
\author{L.~De Sanctis\footnote{Department of Mathematics,
Princeton University, Fine Hall, Washington Road,
Princeton NJ 08544--1000 \ USA \ 
{\tt<lde@math.princeton.edu>}}
\and M.~Isopi\footnote{Dipartimento di Matematica, Universit\`a di Bari,
Via E. Orabona 4, 70125 Bari, Italia \
{\tt<isopi@dm.uniba.it>}, \  {\tt<isopi@cims.nyu.edu>}}
}
\date{July 11, 2001}
\maketitle

\

\leftline{Dedicated to Giovanni Jona-Lasinio on the occasion of his} 
\leftline{seventieth birthday.}

\

\begin{abstract}
Derrida et al. and Sch\"{u}tz and Stinchcombe gave algebraic
formulas for the correlation functions of the partially
asymmetric simple exclusion process. Here we give a fairly general recipe
of how to get these formulas and extend them to the whole
time evolution (starting from the generator of the process),
for a certain class of interacting systems.
We then analyze the algebraic relations obtained 
to show that the matrix approach does not work with some models
such as the voter and the contact processes.
\end{abstract}

\vspace{.5cm}

\noindent{\em Key words and phrases:} simple 
exclusion process, matrix product states,
open system, stationary non equilibrium states.

\newpage

\section{Introduction}
\label{intro}

A few years ago Derrida et al.$^{\cite{derr3}, \cite{derr2}}$
suggested an intriguing ``matrix approach" to the 
one-dimensional Asymmetric Simple 
Exclusion Process (ASEP). This approach
has later been used to treat variants of the 
model$^{\cite{derr4}, \cite{hinrich1}, \cite{hone-pesch}}$,
extended to non steady states and by 
Sch\"{u}tz et al.$^{\cite{schuetz1}, \cite{schuetz2}, \cite{schuetz4}}$, 
used to study 
fluctuations by Derrida et al.$^{\cite{derr5}}$ and the
multispecies case by (among others)
Isaev et al. $^{\cite{ipr}}$.

The main aim of this paper is to study the difficulties that arise 
in potential 
applications of the matrix approach to cases in which
the nearest neighbor interaction or the particle conservation
(both present in the ASEP) are violated. 
Further light on the applicability
of the matrix method is shed by the integrability criterion 
illustrated by Popkov
et al.$^{\cite{schuetz4}}$.

In section \ref{mps} we provide 
a general recipe (using the generator of the process)
to find the algebra of the matrix formalism 
associated to both the steady state and the whole 
dynamics of any one-dimensional interacting system
such that at each step the configuration changes only in two adjacent sites.
A more complete description, with a 
pedagogical aim will be given elsewhere$^{\cite{DI}}$.
In section \ref{beyond} we apply the recipe to some important interacting
systems such as the contact and voter models and show that
the matrix algebra obtained is not useful to treat them.

We will consider only systems in the lattice $\{1,
\ldots , N\}$, this is an intrinsic limitation of the matrix approach. 
The dynamics of an interacting particle system is 
usually defined by giving 
the generator of the process, the general form of which
can be found for instance in Liggett's book$^{\cite{liggett2}}$.

For example, the generator $\Omega$ of 
the ASEP, if particles jump one site to the right (left) with
rate $p$ ($q=1-p$) and
enter the lattice from the left (right) at rate $\alpha$ 
($\delta$) and leave it at rate
$\gamma$ ($\beta$),   
is  defined by
$^{\cite{liggett2}}$:
\begin{multline}\label{liggen}
( \Omega f ) (\tau)  =\\ \!\sum_{x=1}^{N-1} [p\tau(x) (1\!-\tau(x+1))
\! + q\tau(x+1) (1\!-\tau(x))]
[ f( \tau^{x,x+1} ) -\! f( \tau ) ] \\
+[\alpha(1-\tau(1))+\gamma\tau(1)][f(\tau^{1})-f(\tau)] \\ +
[\delta(1-\tau(N))+\beta\tau(N)][f(\tau^{N})-f(\tau)]
\end{multline}
where $\tau=\{\tau(x)\}_{x=1}^N$ is the configuration of the
system, $\tau^{x,y}$ is the
configuration obtained from $\tau$ by exchanging the content of
the sites $x$ and $y$, and $\tau^x$ is the 
configuration obtained from $\tau$ by
changing the content of the $x$-th site.

In the following formulas, $|V\rangle\rangle$ is a 
vector in an (as yet)
unspecified linear space equipped with an inner product, 
$D$ and $E$ linear operators on the same
space, $\langle\langle W|$ is an element of the dual space.
So $\langle\langle W|A|V\rangle\rangle$ is the inner product 
generally written as $({\mathbf W}, A{\mathbf V})$.

The formula of Derrida et al. to write the probability of a given
configuration in the stationary state of the ASEP is$^{\cite{derr1}}$
\begin{equation}\label{derrstst}
P_N(\tau_1,...,\tau_N)=\frac{1}{Z_N}\langle\langle
W|\prod_{j=1}^{N}[\tau_jD+(1-\tau_j)E]|V\rangle\rangle\ ,
\end{equation}
where $D$, $E$, $|V\rangle\rangle$, $\langle\langle W|$ 
are matrices and vectors that satisfy 
\begin{eqnarray}\label{derralg}
(\beta D - \delta E)|V\rangle\rangle  & = & |V\rangle\rangle \ ,\nonumber \\
pDE - qED   & = & D+E \ ,                   \\
\langle\langle W|(\alpha E - \gamma D) & = & \langle\langle W|\nonumber\
\end{eqnarray}
and $Z_N$ is a normalization factor.

One can check these formulas provide a sufficient condition
for the measure to be stationary by observing they
satisfy the recursion relations for the probabilities (first due
to Liggett$^{\cite{liggett3}}$) that relate the probabilities for the
system with $K$ sites to the ones for the system 
with $K-1$ sites$^{\cite{derr1}}$.

\section{From the Generator to Matrix \\ Product States}
\label{mps}

Let us start by re-writing the generator by making use
of a formalism borrowed from quantum mechanics.
For all
$ j=1,...,N$ let us define the Hilbert space   
${\cal H}_j :=span\left\{ \left| 0  \right\rangle _j,\left|
  1 \right\rangle _j\right\} \cong {\mathbb C}^2\ .$
Consider the operators $a^+, a^-, n, m$ defined by:
$a^+|0\rangle  =|1\rangle , a^-|0\rangle  = 0\ ,  
n|0\rangle = 0\ ,  
m =  \mathbb{I}-n\ ,
a^+|1\rangle =0\ ,
a^-|1\rangle =|0\rangle\ ,
n|1\rangle   =|1\rangle$,
where $\mathbb{I}$ is the identity.
Interpreting $|0\rangle$ and $|1\rangle$ as empty 
site and occupied site respectively, the role of $a^+, a^-, n$
as {\em creation, annihilation, number operators} respectively
is rather obvious.
The most immediate choice of an explicit expression for 
the operators and vectors above is
$|0\rangle   = \binom{0}{1}\ ,  |1\rangle   = \binom{1}{0}
a^+  = \begin{pmatrix}
0 & 1 \\     
0 & 0
     \end{pmatrix}, 
 n  = \begin{pmatrix}
    1 & 0 \\
    0 & 0
    \end{pmatrix} , 
        a^-  = \begin{pmatrix}
      0 & 0 \\
      1 & 0
     \end{pmatrix},
   m  = \begin{pmatrix}
      0 & 0 \\
      0 & 1
    \end{pmatrix}
   $.
Now we take the tensor product 
$\mathrm{H}_{N}=\bigotimes_{j=1}^{N}{\cal H}_j$ to describe 
the system on all the $N$ sites. 

If we consider for example the ASEP, in this 
``quantum hamiltonian'' formalism$^{\cite{schuetz3}}$ the
generator is given by
\begin{align}\label{asepham}  
H &=  -\sum_{k}p(a_k^-a_{k+1}^+-n_km_{k+1})+q(a_k^+a_{k+1}^--m_kn_{k+1})+\nonumber\\
&\quad\gamma(a^-_{1} - n_{1}) + \alpha(a^{+}_{1} - m_{1})
   + \beta(a^-_{N} - n_{N}) + \delta(a^{+}_{N} - m_{N})\nonumber\\
   &= h_1^\partial+\sum_{k}h_k+h_N^\partial\ ,
   \end{align}
where the superscript $\partial$ denotes a boundary term and
   $$h_1^\partial=
   \begin{pmatrix}
   \gamma  & -\alpha \\
   -\gamma  &  \alpha
   \end{pmatrix},\
   h_k=\begin{pmatrix}
   0 & 0  & 0  & 0 \\
   0 &  p & -q & 0 \\
   0 & -p & q  & 0 \\
   0 & 0  & 0  & 0
   \end{pmatrix},\ h_N^\partial=
   \begin{pmatrix}
    \beta & -\delta \\
   -\beta &  \delta
   \end{pmatrix}.$$
For any given operator or vector $b$ in the space ${\cal H}_k$ 
we use the notation 
$b_k\equiv \mathbb I\otimes\cdots\otimes\mathbb I\otimes b
\otimes\mathbb I\cdots\otimes\mathbb I$
with $b$ as $k$-th factor.   
Using a different $h_.$, this formulation can be used for any process 
(like the voter and contact, e.g.) such that the occupation
number of each site is either 0 or 1, and such that the dynamics involves 
a couple of neighboring sites at a time (slight generalizations
can be treated as well$^{\cite{hinrich1}\cite{hone-pesch}}$).
The generator in (\ref{asepham}) is the same as in (\ref{liggen})
as can be checked by computing the Dirichlet Form 
for both and verifying that they coincide
(the same holds for processes with different $h$).
It is however easier to look closely at each 
part and see what it does. For instance $a^+a^-$ 
represents a jump to the right and $nm$ takes into account
the complementary event (the particle stays where it is).

We now look for a stationary solution of the master equation 
\begin{equation}\label{master}
   \dot{|P(t)\rangle}=H|P(t)\rangle\ 
\end{equation}
which describes the dynamics of the system by giving the time evolution
of the vector of probabilities of configurations, i.e.
we look for a distribution $|P_s\rangle$ such that
$H|P_s\rangle = 0$. 

In order to show where the general idea can be guessed from, let us
consider again the case of the ASEP, to show$^{\cite{hinrich1}}$ 
that under special conditions (namely 
$(\alpha+\beta+\gamma+\delta)
(\alpha\beta-\gamma\delta)/(\alpha+\delta)(\beta+\gamma)=p-q$)
the stationary state is a product state: 
$|P_s\rangle=\frac{1}{Z_N}\binom{d}{e}^{\otimes N}$ 
 (where
 $d=(\alpha+\delta)/(\alpha\beta-\gamma\delta)$ ,
 $e=(\beta+\gamma)/(\alpha\beta-\gamma\delta)$ and
 the normalization constant is clearly
  $Z_N=(e+d)^N$).
 To prove that $H|P_s\rangle=0$, one should first check
 \begin{equation}\label{tmr}
  h_i\left[\binom{d}{e}\otimes\binom{d}{e}\right]= 
\binom{d}{e}\otimes\binom{-1}{1}
 -\binom{-1}{1}\otimes\binom{d}{e}\ .\end{equation} 
This makes the sum through which $H$ is defined telescopic
(recall that we are omitting the factors of the tensor product
on which the operators act trivially as the identity), 
and since
$$
  h_1^\partial\binom{d}{e}=  \binom{-1}{1}\ ,\
   h_N^\partial\binom{d}{e}= -\binom{-1}{1}\ ,
$$
the cancellation of the first term with the last is assured by
the boundary terms. 

In other words, $H|P_s\rangle = 0$ would be solved for instance
if we had zero for all $i$ in the r.h.s. of (\ref{tmr}); but
this is too restrictive, so we look for the first non trivial possibility:
instead of zero, we impose a ``telescopic term". This is inspired by
the dynamics, that acts with the same $h_.$ on all couple of 
adjacent sites, so the generator acts twice on each site.
We will now try to make the above approach work 
for non-product states by imposing a similar 
telescopic property. 
The idea is to move into a richer context,
substituting the numbers 1, $e$,
$d$ appearing in (\ref{tmr}) with some time-dependent 
operators (non commuting
and acting on an auxiliary space of generally infinite dimension) 
$S$, $E$, $D$ to be determined, aiming to get the 
weights of each possible configuration
through a bracket with a couple of vectors $\langle\langle W|$ 
and 
$|V \rangle\rangle$ to be introduced in the same space.
For instance for a system consisting of a single site we would impose
$
\langle\langle W|\binom{D}{E}|V \rangle\rangle 
=\binom{\langle\langle W|D|V \rangle\rangle}
{\langle\langle W|E|V \rangle\rangle} 
= \binom{d}{e}
$
  and clearly, in the case of a product measure,
$
   \langle\langle W|\binom{D}{E}^{\otimes N}|V \rangle\rangle =
 \binom{d}{e}^{\otimes N}
$.
We can also write
$
  H \langle\langle W|\binom{D}{E}^{\otimes N}|V \rangle\rangle =
   \langle\langle W|H \binom{D}{E}^{\otimes N}|V \rangle\rangle
$. 

Let us now write
 $
   |P \rangle=\frac{1}{Z_N} \langle\langle
   W|\binom{D}{E}^{\otimes N}|V \rangle\rangle
$ for the probability vector 
 and plug it into the master equation (\ref{master}).
 Clearly $Z_N=\langle\langle
 W|C^N|V \rangle\rangle$, with $C=D+E$, that does not
 depends on time by conservation of probability.

It is easy to show that the master equation (\ref{master}) is satisfied
if the following equalities hold 
(thanks to the same telescopic cancellation mechanism we used for 
the product state)
  \begin{align}\label{algebra}
   & \left(\frac{1}{2}\frac{d}{dt}_.+h_.\right)
   \binom{D}{E}\otimes\binom{D}{E} =\nonumber \\
   & \quad \binom{D}{E}\otimes\binom{-S}{S}-
      \binom{-S}{S}\otimes \binom{D}{E} \ ,  \\
  & \langle\langle W|\left[(\frac{1}{2}\frac{d}{dt}+h_1^\partial)
   \binom{D}{E}-\binom{-S}{S}\right]=0\ ,\nonumber\\
   & \left[(\frac{1}{2}\frac{d}{dt}+h_L^\partial)
   \binom{D}{E}+\binom{-S}{S}\right] |V\rangle\rangle =0\nonumber\ .
   \end{align}
These are the relations of the matrix algebra of the process.
If we chose for example the $h_.$ of the ASEP, these equations take the 
explicit form of the algebra found by Stinchcombe
and Sch\"{u}tz$^{\cite{schuetz1}, \cite{schuetz2}}$
that includes as a special 
case the stationary one (\ref{derralg}) of Derrida et al. 
(taking $S=\mathbb{I}$ and putting all the time derivatives equal to zero).
With this procedure we can exhibit an algebra 
for all the models with a dynamics involving
only a couple of neighboring sites at a time$^{\cite{schuetz3}, \cite{schuetz7}}$
(see the same works  for a classification of the models
with different $h_.$).
If one found an explicit expression for all the operators and vectors, the model 
could in principle be solved exactly (provided the algebra is not empty). 
Unfortunately, this is in general very difficult
to accomplish (a purely algebraic treatment can also 
be used$^{\cite{schuetz4}}$). 
In the case of the ASEP, thanks to the preservation 
of the number of particles in the bulk 
dynamics, the local generator $h_.$ has a block form,
with zero entries in the first and last row and column.
This special form of $h_.$ is such that 
in stationary conditions the four equations 
(\ref{algebra}) collapse to just one: (\ref{derralg}). 
But this great simplification may not occur for 
different models.
In many cases the algebra
can be empty (or too complicated to deal with), 
as we are going to show 
for the contact and voter models.
We can say that the method works for the processes, 
such as the ASEP, the probability measures of which are either
product, or a generalization that we can classify as 
``matrix product measures".  If one distinguished only 
between product and non-product states, the choice would be 
in general only 
between a numerical tensor product and a convex combination
of  as many such products as the cardinality of the configuration
space. If the states of a process are matrix product, one can chose
to deal again with a single tensor product, thanks to the richer
nature of the entries, matrices instead of numbers.

Algebras defined by conditions like (\ref{algebra}) are called 
\emph{Diffusion Algebras} $^{\cite{ipr}}$.

 \section{The matrix Approach Beyond \\ Simple Exclusion}
\label{beyond}

 \subsection{Exclusion process with double jumps}

The method to write the matrix algebra of the process
can also be extended to the 
case of dynamics not limited
to neighboring sites, such as for instance 
the exclusion process with jumps
of length two permitted. The generator, in the case of 
symmetric dynamics, is (up to boundary terms):
\begin{align*}\begin{split}
   H = &
   -\sum_{k}(a_k^-a_{k+1}^+-n_km_{k+1})+(a_k^+a_{k+1}^--m_kn_{k+1})+\\
   & (a_k^-a_{k+2}^+-n_km_{k+2})+(a_k^+a_{k+2}^--m_kn_{k+2})=\sum_{k}h_k,
   \end{split}\end{align*}
$$
h_.=\begin{pmatrix}
     0 & 0 & 0 & 0 & 0 & 0 & 0 & 0 \\
     0 & 1 & 0 & 0 & -1 & 0 & 0 & 0\\ 
     0 & 0 & 1 & 0 & -1 & 0 & 0 & 0\\
     0 & 0 & 0 &  2 & 0 & -1 & -1 & 0\\
     0 & -1 & -1 &  0 & 2 & 0 & 0 & 0\\
     0 & 0 & 0 &  -1 & 0 & 1 & 0 & 0\\
     0 & 0 & 0 &  -1 & 0 & 0 & 1 & 0\\
     0 & 0 & 0 & 0 & 0 & 0 & 0 & 0\
   \end{pmatrix}
$$
For this system we impose the telescopic property to solve
the master equation in the following way:
\begin{align*}
&\left(\frac{1}{3}\frac{d}{dt}_.+h_.\right)\binom{D}{E}\otimes\binom{D}{E}\otimes\binom{D}{E}=
\binom{D}{E}\otimes\binom{-S}{S}\otimes\binom{D}{E}-\\
 &\qquad    2\binom{-S}{S}\otimes \binom{D}{E}\otimes\binom{D}{E}+
\binom{D}{E}\otimes\binom{D}{E}\otimes\binom{-S}{S}
\end{align*}
which is the same as
\begin{equation*}
\frac{1}{3}(2\dot{D}D^2+D\dot{D}D+D^2\dot{D})+ 0                              
= -DSD+2SD^2-D^2S
\end{equation*}
\begin{equation*}
\frac{1}{3}(2\dot{D}DE+D\dot{D}E+D^2\dot{E})+ D^2E-ED^2             
= -DSE+2SDE+D^2S   
\end{equation*}
\begin{equation*}
\begin{split}
\frac{1}{3}(2\dot{D}ED+D\dot{E}D+DE\dot{D})+ DED-ED^2                
=\\ DSD+2SED-DES
   \end{split}
\end{equation*}
\begin{equation*}
\begin{split}
\frac{1}{3}(2\dot{D}E^2+D\dot{E}E+DE\dot{E})+ 
2DE^2-EDE-E^2D  = \\ DSE+2SE^2+DES
   \end{split}
\end{equation*}
\begin{equation*}
\begin{split}
\frac{1}{3}(2\dot{E}D^2+E\dot{D}D+ED\dot{D}) 
-D^2E-DED+2ED^2 =\\ -ESD-2SD^2-EDS
   \end{split}
\end{equation*}
\begin{equation*}
\begin{split}
\frac{1}{3}(2\dot{E}DE+E\dot{D}E+ED\dot{E})-DE^2-EDE                    
=\\ -ESE-2SDE+EDS
   \end{split}
\end{equation*}
\begin{equation*}
\frac{1}{3}(2\dot{E}ED+E\dot{E}D+E^2\dot{D})-DE^2+E^2D               
=ESD-2SED-E^2S
\end{equation*}
\begin{equation*}
\frac{1}{3}(2\dot{E}E^2+E\dot{E}E+E^2\dot{E})- 0  =ESE-2SE^2+E^2S
\end{equation*}
These relations define now a cubic algebra, as opposed to a quadratic one,
which is therefore not a Diffusion Algebra in the sense of $^{\cite{ipr}}$.
Unfortunately algebras of degree higher than two are very difficult
to handle (see e.g. Vershik$^{\cite{vershik}}$).
However algebras of degree higher than two appear e.g. in
$^{\cite{schuetz4}}$.

\subsection{Voter and Contact Models}

For a description of the voter and contact models
see Liggett$^{\cite{liggett2}}$.
It is easy to see that the local 
generator for the voter model can be written 
in the form of the r.h.s. of (\ref{asepham})
with
   $$
  h_.=\begin{pmatrix}
     0 & -1 & -1 & 0 \\
     0 & 2  &  0  & 0 \\
     0 & 0  &  2  & 0 \\
     0 & -1 & -1 & 0 \
   \end{pmatrix} 
  $$
   and 
$$
h_1=\begin{pmatrix}
     0 & -\lambda \\
     0 & \lambda \
   \end{pmatrix}\ ,\ h_N=\begin{pmatrix}
     \mu & 0 \\
     -\mu & 0 \
   \end{pmatrix}
$$
where $\lambda$ and $\mu$ are the rates for opinion changing in
the boundary sites. Notice that there are non zero entries in the 
first and last row.
It is easy to compute that
   $$
   h_.\binom{D}{E}\otimes\binom{D}{E}=\left(\begin{array}{c}
     -\{D,E\} \\
     2DE \\
     2ED \\
     -\{D,E\} \
   \end{array}\right)
   $$
   and so we can conclude that the algebra and its stationary limit
   are given by
   \begin{align*}
   \frac{1}{2}(\dot{D}D+D\dot{D})-\{ D,E \}
   &= [S,D] \longrightarrow \{D,E\}=0 \\
   \frac{1}{2}(\dot{D}E+D\dot{E})+2DE &= SE+DS \longrightarrow
   2DE=C \\
   \frac{1}{2}(\dot{E}D+E\dot{D})+2ED &= -(SD+ES) \longrightarrow
   2ED=-C \\ 
  \frac{1}{2}(\dot{E}E+E\dot{E})-\{D,E\}
   &= [E,S]\longrightarrow \{D,E\}=0
   \end{align*} 
Hence in stationary conditions $$[D,E]=C\equiv D+E ,\{D,E\} = 0 ,
  \mu D|V\rangle=|V\rangle  , \langle W|\lambda E
   =\langle W|.
$$
Notice that the relations are similar to the ones of the ASEP, but
there is an additional condition: $D$ and $E$ anticommute.

The local generator of the contact model is
   $$
   h_.=\begin{pmatrix}
    0 & -\alpha & -\alpha & 0 \\
     0 & \alpha+\beta & 0 & -\alpha \\
     0 & 0 & \alpha+\beta & -\alpha \\ 
    0 & -\beta & -\beta & 2\alpha \
   \end{pmatrix}
   $$
   so that
   $$
   h\binom{D}{E}\otimes\binom{D}{E}=\left(\begin{array}{c}
     -\alpha\{D,E\} \\
     (\alpha+\beta)DE-\alpha E^2 \\
     (\alpha+\beta)ED-\alpha E^2 \\
     -\beta\{D,E\}+2\alpha E^2 \
   \end{array}\right)
   $$
   and so we can conclude that the algebra 
   is given by   
   \begin{align*}
   \frac{1}{2}(\dot{D}D+D\dot{D})-\alpha\{D,E\}
   &= [S,D] \\
   \frac{1}{2}(\dot{D}E+D\dot{E})+(\alpha+\beta)DE-\alpha E^2
   &= SE+DS \\
   \frac{1}{2}(\dot{E}D+E\dot{D})+(\alpha+\beta)ED-\alpha E^2
   &= -(SD+ES) \\
   \frac{1}{2}(\dot{E}E+E\dot{E})-\beta\{D,E\}+2\alpha E^2
   &= [E,S]
   \end{align*}
   so that in stationary conditions $E^2=0,\ [D,E]=C\  , \{D,E\} = 0$
   if we assume $\alpha=\beta=1$.

Clearly these relations define a subalgebra 
of the one for the voter model.

\begin{theorem}
In stationary conditions, the algebra of the voter model is empty 
(and {\em a fortiori} so is the one of the contact process and so
are the ones for the whole time evolution).
\end{theorem}
{\bf Proof }
 The algebra is
   $$
    DE=(D+E)/2\ ,\ ED=-(D+E)/2\ , $$
    $$
    DE=-ED\ ,\
    D|V\rangle=\mu|V\rangle\ ,\
    \langle W|E=\langle W|\lambda\ .
   $$
   If
   $$
    \vartheta_.=D,\ E$$ we get, from the first two conditions
   $$ \langle W|\prod_{k=1}^N\vartheta_k|V\rangle=
    \langle W|[P(D)+Q(E)]|V\rangle=
    [P(1/\mu)+Q(1/\lambda)]\langle W|V\rangle
   $$
with some polynomials $P$ and $Q$; but the third condition 
(anticommutation) also implies
   $$
    \langle W|\prod_{k=1}^N\vartheta_k|V\rangle=(\pm)
    \langle W|E^mD^n|V\rangle=(\pm)(1/\lambda)^m(1/\mu)^n
    \langle W|V\rangle
   $$
  where $m+n=N$. The two expressions cannot be equal for all 
values of $\lambda$ and $\mu$. $\Box$

This shows that, following the recipe of section (\ref{mps}), 
we cannot use the matrix approach. However, the l.h.s of 
(\ref{algebra}) reflects directly the dynamics of the process
and does not depend on the matrix formalism, but the telescopic r.h.s.
is only inspired by the nearest neighbor nature of the dynamics
and it is more ``artificial".
In other words, if another way to solve the master equation were developed,
some kind of matrix approach could still be productive also for
those models that cannot be treated with the current matrix approach
illustrated in this paper.

 \section*{Acknowledgments}
The authors thank G. M. Sch\"{u}tz for useful discussions
and making available reference$^{\cite{schuetz7}}$ before publication;
and an anonymous referees for pointing out 
references$^{\cite{schuetz4}, \cite{ipr}}$.
Useful suggestions from two referees, as well as the editor, led to an
improved presentation of our results.
 
L.D.S. would like to thank G. Jona-Lasinio for his encouragement to
complete this work.

\end{document}